\documentstyle[12pt,twoside,fleqn,espcrc1,epsf]{article}
\newcommand{\ZPC}[3]{{Z. Phys. C} {#1} (#2) {#3}.}
\newcommand{\PLB}[3]{{ Phys.~Lett. B} {#1} (#2) #3.}
\newcommand{\HIP}[3]{{ Heavy Ion Physics} {#1} (#2) #3.}
\newcommand{\NPA}[3]{{ Nucl. Phys. A} {#1} (#2) #3.}

\newcommand{\JPG}[3]{{ J. Phys. G} {#1} (#2) #3.}

\newcommand{\AmS}{{\protect\the\textfont2
  A\kern-.1667em\lower.5ex\hbox{M}\kern-.125emS}}

\title{Canonical Strangeness Enhancement\thanks{Work supported by 
BMBF and DFG.}}

\author{\underline{J. Sollfrank},\address{Institut f\"ur Theoretische Physik,
                     Universit\"at Regensburg,\\ D--93040 Regensburg, Germany}
        F. Becattini,\address{INFN Sezione di Firenze,
                    Largo E. Fermi 2,\\ I--50125 Firenze, Italy}
        K. Redlich,\address{Institute for Theoretical Physics,
                   University of Wroclaw,\\ PL--50204 Wroclaw, Poland}
        and
        H. Satz\address{Fakult\"at f\"ur Physik,
                   Universit\"at Bielefeld,\\ D--33501 Bielefeld, Germany}}

\begin{document}

\maketitle

\begin{abstract}
 According to recent experimental data and theoretical developments we 
 discuss three distinct topics related to strangeness enhancement in nuclear 
 reactions. We investigate the compatibility of multi-strange particle ratios
 measured in a restricted phase space with thermal model parameters 
 extracted recently in $4\pi$. We study the canonical suppression as 
 a possible reason for the observed strangeness enhancement and
 argue that a connection between QGP formation and the undersaturation of
 strangeness is not excluded.
\end{abstract}

\section{Multi-strange hadrons from thermal models}
Particle production in heavy ion collisions at SPS energies has been 
recently analyzed in \cite{Becattini97b}. The analysis 
includes particle abundances and ratios measured or extrapolated to 
full momentum space. The use of $4\pi$ data makes the analysis 
independent of the particle momentum distributions which are quite different 
from the ones of a static fireball.
\begin{table}[hbt]
\caption{Comparison of (multi-)strange particle ratios with the
thermal model. The kinematic cuts applied are 
$2.5 < y < 3.0$, $1.2 < p_T< 3.0 $ (S+S \cite{wa94}), 
$2.3 < y < 3.0$, $1.2 < p_T< 3.0 $ (S+W \cite{wa85}), and
$2.4 < y < 3.4$, $0.7 < p_T $ (Pb+Pb \cite{wa97}). }
\label{tab1}
\begin{tabular*}{\textwidth}{@{}l|@{\extracolsep{\fill}}cc|cc}
\hline
&\multicolumn{2}{c|}{WA94 S  + S 200 A GeV}&
\multicolumn{2}{c}{WA85 S  + W 200 A GeV}\\
ratio & experiment \cite{wa94}& thermal model & 
experiment \cite{wa85}& thermal model\\
\hline
$\overline{\Lambda} / \Lambda$ & 
      $0.23\pm0.01$ & $0.176 \pm 0.017$ &
      $0.196\pm0.011$ & $0.156 \pm 0.015$ \\
$\overline{\Xi} / \Xi$ & 
      $0.55\pm0.07$ & $0.327 \pm 0.046$ &
      $0.47\pm0.06$ & $0.303 \pm 0.042$ \\
$\Xi / \Lambda$ & 
      $0.09\pm0.01$ & $0.124 \pm 0.009$ &
      $0.097\pm0.006$ & $0.120 \pm 0.009$ \\
$\overline{\Xi} / \overline{\Lambda}$ & 
      $0.21\pm0.02$ & $0.231 \pm 0.023$ &
      $0.23\pm0.02$ & $0.228 \pm 0.022$ \\
\hline
&\multicolumn{2}{c|}{WA97 Pb + Pb 158 A GeV} & &\\ 
ratio & experiment \cite{wa97}& thermal model & & \\ 
\hline
$\overline{\Lambda} / \Lambda$ & 
      $0.124\pm0.013$ & $0.179 \pm 0.017$ & & \\
$\overline{\Xi} / \Xi$ & 
      $0.255\pm0.025$ &  $0.328 \pm 0.045$ & & \\
$\overline{\Omega} / \Omega$ & 
      $0.38\pm0.10$ &  $0.61 \pm 0.12$ & & \\
$\Xi / \Lambda$ &   $0.099\pm0.008$ &  $0.100 \pm 0.007$ & & \\
$\overline{\Xi} / \overline{\Lambda}$ & 
      $0.203\pm0.024$ &  $0.184 \pm 0.018$ & & \\
$\Omega / \Xi$ & 
      $0.192\pm0.024$ & $0.126 \pm 0.011$ & & \\
\end{tabular*}
\end{table}
In order to test the significance of the results derived in \cite{Becattini97b}
we investigate its compatibility with multi-strange baryon ratios
omitted in \cite{Becattini97b} due to their measurements
in limited phase space. In order to cut the thermal model prediction 
for these ratios to the experimental acceptance we define an acceptance 
correction factor $C_r$ for each considered particle ratio $r$ by the ratio 
\begin{equation} \label{kincor}
C_r = \frac{N^{\rm acc}}{N^{4\pi}}\; .
\end{equation} 
In order to calculate $C_r$ we use the momentum distributions of 
$\Lambda$'s, $\Xi$'s and $\Omega$'s as they result from the
hydrodynamical evolution of the corresponding collision \cite{Sollfrank98}.
It has been shown \cite{Sollfrank98} that this evolution scenario
describes quite well the momentum distributions of various hadron
species.

The analysis presented in \cite{Becattini97b} shows that the intensive thermal 
parameters are insensitive to the nuclear collision size. At SPS energies 
hadronization can then be characterized by a universal set of parameters
$T = 180 \pm 10$ MeV, $\gamma_s = 0.7 \pm 0.05$
and $\mu_{\rm B}/T = 1.25 \pm 0.1$. Taking these values as input for
a thermal model calculation and multiplying the resulting ratios with $C_r$
of Eq. (\ref{kincor}) we get the results shown in Table \ref{tab1}. 
They are in agreement with the measurements of WA94 \cite{wa94}, 
WA85 \cite{wa85} and WA97 \cite{wa97} within 2--3 standard deviations. 
Taking into account the model uncertainties in the acceptance corrections 
we find that also the multi-strange baryons are compatible with an 
undersaturation of strangeness of order $\gamma_s = 0.7 \pm 0.05$.

\section{Canonical strangeness enhancement}

\begin{figure}[htb]
\begin{minipage}[t]{79mm}
\epsfxsize 79mm \epsfbox{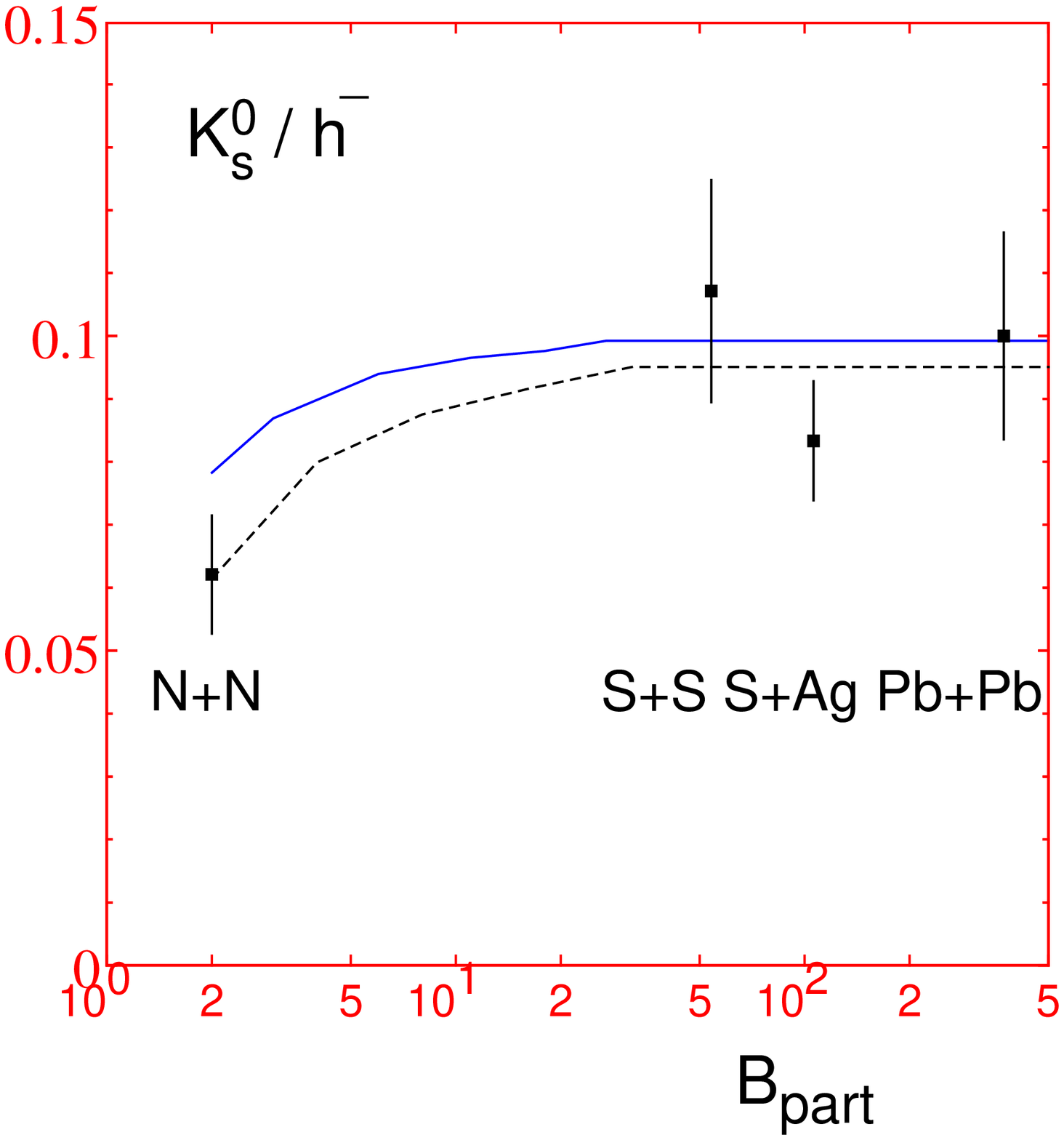}
\end{minipage}
\hspace{\fill}
\begin{minipage}[t]{79mm}
\epsfxsize 79mm \epsfbox{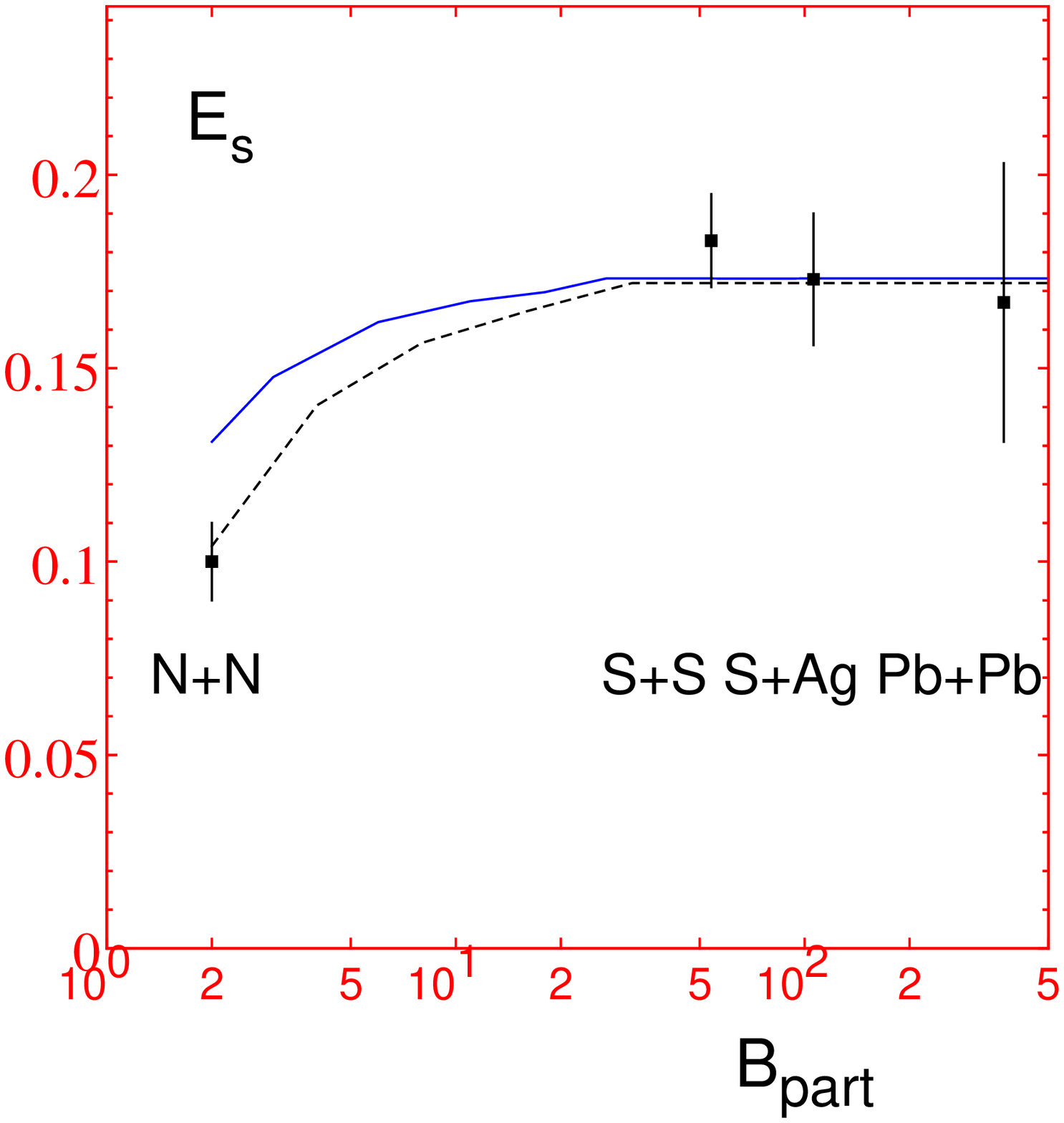}

\end{minipage}
\caption{ $K_s^0/h^-$ and $E_S$ as function of 
$B_{\rm part}$ in a thermal model with $T = 180$ MeV and $\gamma_s = 0.7$.
The volume $V$ is $V = 5$ fm$^3 \times B_{\rm part}$ (solid line) and
$V = 2.5$ fm$^3 \times B_{\rm part}$ (dashed line). The data for $E_S$ are
taken from the compilation in \cite{Gazdzicki96} and for $K_s^0/h^-$ from
the compilation in \cite{Becattini97b} and \cite{Gazdzicki91} (N+N, i.e.
isospin averaged nucleon nucleon collision).}
\label{esfig}
\end{figure}

Various experiments report an enhancement of strange to non-strange
particle ratios going from p+p to A+A collisions \cite{wa97,Gazdzicki96}.
On the other hand, in a thermal model the strange hadrons are 
suppressed by exact conservation of quantum numbers when going from large
volumes (A+A collisions) to small ones (p+p collisions).
The immediate question is whether the canonical
suppression can account for the observed effect. Since the $\gamma_s$
values for p+p collisions $\gamma_s = 0.5 \pm 0.05$ \cite{Becattini97a} and 
for A+A collisions $\gamma_s = 0.7 \pm 0.05$ \cite{Becattini97b} are clearly 
different while the hadronization temperature $T = 180 \pm 10$ MeV is 
rather universal the volume increase is not the only effect which
accounts for the total strangeness enhancement. This is shown 
quantitatively in Figure \ref{esfig} where the ratios $K_s^0/h^-$ and 
$E_S = (\langle \Lambda \rangle + \langle K \rangle)/ \langle \pi \rangle$
are shown as function of participating
baryons $B_{\rm part}$. Taking the volume per participating baryon as
$V_0 = V/B_{\rm part}  = 5$ fm$^3$, a value which is also in agreement with 
the total abundances, then we find that the canonical suppression 
accounts only for half of the observed effect.

On the other hand, if we take only one half of the original 
volume, i.e. $V_0/B_{\rm part}  = 2.5$ fm$^3$, then the strangeness
enhancement in particle ratios can be accounted for by the canonical 
suppression alone (see Figure \ref{esfig}, dashed line). One gets 
$\gamma_s \approx 0.7$ for p+p collisions as also seen in 
$e^+ + e^-$ \cite{Becattini96} and in A+A \cite{Becattini97b} collisions.
This would imply universality of $\gamma_s$. In order to get the 
proper values of the total particle abundances in such a scenario at 
least two small fireballs are required which have to hadronize 
independently. E.g. for p+p collisions a model with two fireballs 
associated to the leading protons may be constructed.
However, this suggestion suffers from a serious difficulty connected 
with the measured $\phi$ yield. The $\phi$-meson is not canonically 
suppressed because it has zero charges. For $\gamma_s \approx 0.7$ in 
p+p collisions the measured $\phi$ abundance deviates from a thermal fit by
$12\sigma$, increasing even more the already troublesome deviation of 
$4\sigma$ for $\gamma_s = 0.5$ \cite{Becattini97a}.

\section{Saturation of $\gamma_s$}
In order to understand the systematics of hadron production
it is important to look at the $A$-scaling of $\gamma_s$.
It was found \cite{Becattini97b} that the chemical freeze-out in 
nuclear collisions at CERN-SPS appears with rather constant
$\gamma_s = 0.7 \pm 0.05$.  This result suggests that
undersaturation of strangeness is independent of the system's
size and the life time of the fireball.
\begin{figure}[htb]

\begin{minipage}[b]{75mm}
\epsfxsize 75mm \epsfbox{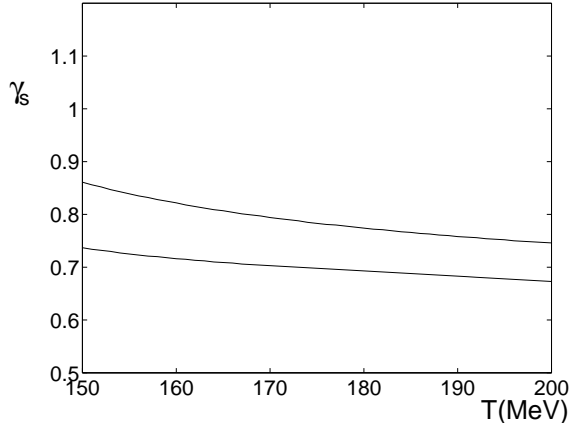}
\end{minipage}
\hspace{\fill}
\begin{minipage}[b]{64mm}
\caption{$\gamma_s$ of the hadron gas which gives the same strange
quark multiplicity per entropy as a fully equilibrated
quark-gluon plasma at the same temperature. The net baryon number was
set to zero. The upper
line is for a strange quark mass of $m_s = 130$ MeV, 
the lower line for $m_s = 190$ MeV.
}
\label{fig2}
\end{minipage}
\end{figure}
We now present an argument \cite{Gazdzicki97} for a possible explanation 
of the apparent A-independence of $\gamma_s$ in nuclear collisions.
Let us consider the ratio 
$r_S = \langle s + \overline{s} \rangle/S\;$
of the total number of strange quarks $\langle s + \overline{s} \rangle$ 
over total entropy $S$. Since the ratio $r^{\rm QGP}_S / r^{\rm HG}_S$
is of the same order as the afore mentioned undersaturation of 
strangeness it is conceivable that at high bombarding
energies the value of $\gamma_s$ could be determined by
$\gamma_s \approx r^{\rm QGP}_S/r^{\rm HG}_S.\; $
Since $r^{\rm HG}_S$ depends on $\gamma_s$ in a non-linear way we
solve the implicit equation \cite{Redlich98}
\begin{equation}\label{implicit}
r_S^{\rm QGP}(T,S/A) = r_S^{\rm HG}(T,S/A,\gamma_s) 
\end{equation}  
for $\gamma_s$. The formulation of the hadron gas thermodynamics is
done as in \cite{Becattini97b} while for the QGP an ideal non-interacting
parton gas with two massless flavours and a massive strange quark with
$m_s = 160 \pm 30$ MeV \cite{Jamin97} is used. For zero baryon density 
we get the results shown in Figure \ref{fig2}. In the temperature 
range expected for the deconfinement transition the resulting $\gamma_s$
is of the same order as the one extracted in the chemical analysis 
for particle abundances \cite{Becattini97b}. For finite baryon densities 
the results do not change \cite{Redlich98} qualitatively. 

The hadronization of gluons from a QGP will naturally also feed into the 
strange sector. If this branching of gluons is large the whole idea can 
only be rescued by assuming that in the hadronization process a similar 
increase in entropy is present. Since these processes lie in the 
non-perturbative domain of QCD a quantitative treatment is out of sight. 
The principal possibility, however, that the observed undersaturation of 
strangeness might be due to the hadronization of a partonic gas cannot 
be excluded.

\end{document}